# CEST MR fingerprinting (CEST-MRF) for Brain Tumor Quantification Using EPI Readout and Deep Learning Reconstruction

Ouri Cohen*[1], Victoria Y. Yu[1], Kathryn R. Tringale[2], Robert J. Young[3], Or Perlman[4], Christian T. Farrar[4], Ricardo Otazo[1,3]

[1]Department of Medical Physics, Memorial Sloan Kettering Cancer Center, New York, NY, USA

[2]Department of Radiation Oncology, Memorial Sloan Kettering Cancer Center, New York, NY, USA

[3]Department of Radiology, Memorial Sloan Kettering Cancer Center, New York, NY, USA

[4]Athinoula A. Martinos Center for Biomedical Imaging, Department of Radiology, Massachusetts General Hospital and Harvard Medical School, Charlestown, MA, USA

**Correspondence to:** Ouri Cohen, Memorial Sloan Kettering Cancer Center, 320 East 61st St, New York, NY, 10025, USA.

**Research Article**

## ABSTRACT

**Purpose:** To develop a clinical CEST MR fingerprinting (CEST-MRF) method for brain tumor quantification using EPI acquisition and deep learning reconstruction.

**Methods:** A CEST-MRF pulse sequence originally designed for animal imaging was modified to conform to hardware limits on clinical scanners while keeping scan time $\leq 2$ minutes. Quantitative MRF reconstruction was performed using a deep reconstruction network (DRONE) to yield the water relaxation and chemical exchange parameters. The feasibility of the 6 parameter DRONE reconstruction was tested in simulations in a digital brain phantom. A healthy subject was scanned with the CEST-MRF sequence, conventional MRF and CEST sequences for comparison. Reproducibility was assessed via test-retest experiments and the concordance correlation coefficient (CCC) calculated for white matter (WM) and grey matter (GM). The clinical utility of CEST-MRF was demonstrated in 4 patients with brain metastases in comparison to standard clinical imaging sequences. Tumors were segmented into edema, solid core and necrotic core regions and the CEST-MRF values compared to the contra-lateral side.

**Results:** The DRONE reconstruction of the digital phantom yielded a normalized RMS error of $\leq 7\%$ for all parameters. The CEST-MRF parameters were in good agreement with those from conventional MRF and CEST sequences and previous studies. The mean CCC for all 6 parameters was $0.98 \pm 0.01$ in WM and $0.98 \pm 0.02$ in GM. The CEST-MRF values in nearly all tumor regions were significantly different ($P=0.05$) from each other and the contra-lateral side.

**Conclusion:** Combination of EPI readout and deep learning reconstruction enabled fast, accurate and reproducible CEST-MRF in brain tumors.

# 1. Introduction

CEST MRI uses frequency selective radiofrequency pulses to saturate the magnetization of labile protons on proteins and metabolites [1]. The saturated protons exchange with the unsaturated water protons and lead to a measurable reduction in the water MRI signal. The CEST contrast is attractive since it is sensitive to metabolite concentrations with higher spatial resolution (~1 mm) and shorter scan times (~5 min) than MRS [2]. Moreover, the measured CEST signal depends, inter alia, on the chemical exchange rate, which is pH sensitive. Because many pathologies, including cancer, are characterized by tissue hypoxia leading to an acidic microenvironment, pH is a potentially valuable metabolic biomarker [3]–[6]. In cancer imaging, the CEST contrast has been used to distinguish pseudoprogression and radiation necrosis from true progression in brain tumors [7]–[9], quantify tumor extracellular pH [3], [10], evaluate the grading and cellularity of gliomas [11] and monitor early effects of radiation therapy [12].

Although preclinical and early clinical CEST studies have demonstrated the potential utility of CEST methods for assessing disease pathologies, disease progression and therapeutic response, they have not been widely adopted for clinical use due to several challenges encountered in clinical translation. Specifically, the qualitative nature of the CEST contrast, the relatively long image acquisition times, and the complicated data processing have all hindered clinical translation. To overcome these challenges, MR fingerprinting (MRF)-based CEST was recently introduced [13], [14] and demonstrated in vivo for rat brain experiments on a preclinical 4.7T scanner [15]. The CEST-MRF sequence has the following advantages over conventional CEST: tissue maps are fully quantitative, the acquisition time is short (less than 2 minutes) and the data analysis is greatly simplified because only a single resonance frequency offset is excited with saturation pulses of varying powers (instead of the full Z-spectrum) which also reduces the sensitivity to B0 variations due to the use of a normalized signal. The accuracy of the tissue parameter maps was validated in phantoms and compared to reference methods from the literature [16]–[19].

The diagnostic potential of CEST-MRF for pathologies is considerable since the resulting parametric maps reflect different biophysical processes, and their combination provides a comprehensive picture of complex pathologies, like brain tumors, where multiple parameters change simultaneously. However, to realize the clinical potential of CEST-MRF, the pulse sequence must be adapted to clinical scanners. This is challenging because preclinical scanners have significantly different RF and gradient amplifier capabilities and limitations compared to clinical scanners. For example, patient imaging is bound by strict limits on specific absorption rate and peripheral nerve stimulation. These hardware and software differences require careful consideration when adapting a sequence for clinical use. An additional challenge inherent to CEST-MRF is the large number of tissue parameters quantified by the sequence. In conventional MRF, tissue quantification is achieved by pattern-matching the measured signal to a pre-computed database of signal magnetizations [20]. However, the exponential growth in dictionary size for multiple tissue parameters renders this approach impractical for CEST-MRF. Some groups have reported the use of a MRF framework with a two-pool model in healthy subjects to quantify the water and semi-solid parameters but not the pH-sensitive amide exchange [21], [22]. Others have proposed to sequentially quantify the semi-solid and CEST parameters using dictionary matching reconstruction necessitating two separate scans and a potentially lengthy reconstruction process [23].

The aim of this work is to enable CEST-MRF scans of patients with brain metastases by development of acquisition and quantification methods suitable for clinical scanners and workflow. Specifically, a combination of MRF with EPI readout and deep learning-based quantification is proposed. The proposed method is tested in simulations and a healthy volunteer and validated in vivo against conventional CEST imaging. The clinical utility is demonstrated in patients with metastatic brain cancer.

## 2. Methods

### 2.1. Pulse sequence

Figure 1A shows the proposed pulse sequence for one temporal point of the acquisition schedule. The following temporal points of the acquisition schedule use the same pulse sequence but with different CEST saturation encoding (Figure 1B). The initial pulse train saturates the solute protons and is composed of 160 non-selective, Gaussian-shaped, 16 ms sub-pulses applied with a 100% duty cycle for a total pulse train duration ($T_{sat}$) of 2560 ms. The saturation pulse train power ($B1_{sat}$) was varied according to a pre-determined schedule as described in section 2.1.1. The resonance frequency of the RF pulses was set to the chemical shift of the amide protons (3.5 ppm) in this study but protons of other moieties (amine, hydroxyl etc.) are easily probed by setting the resonance frequency to the appropriate chemical shift. The saturated protons chemically exchange with the unsaturated bulk water protons which leads to a reduction in the water signal which can be measured by excitation with an on-resonance RF pulse with flip angle (FA) set according to the MRF acquisition schedule. The magnetization is then sampled with an EPI readout and the acquisition repeated, following a repetition delay (TR), for each point in the acquisition schedule.

### 2.1.1. MRF acquisition schedule

To ensure a differential signal evolution for different tissues and facilitate their quantification, the acquisition parameters must be varied for each schedule point. Although simultaneous variation of multiple acquisition parameters can improve discrimination and reduce acquisition times [24] the choice of a good schedule is a challenging problem [25], and will be explored in future studies. For simplicity, in this work $B1_{sat}$ was randomly varied for 30 schedule points with powers in the range 0-4µT (Figure 1B), which was found to yield accurate parameters in prior preclinical in vivo studies [15]. The variable saturation gives rise to a differential tissue evolution for different tissue types (Figure 1C). All other acquisition parameters (FA, TR, $T_{sat}$) were kept constant with FA set to 90°, TR set to 3500 ms and $T_{sat}$ set to 2560 ms.

## 2.2. Deep learning-based tissue parameter quantification

In conventional MRF, tissue parameters are quantified by matching the measured signal to a pre-computed dictionary of signal magnetizations. Because of the large number of tissue parameters (dictionary dimensions) and the exponential growth of the dictionary, this approach is infeasible for CEST-MRF. We have previously demonstrated the use of a model-trained neural network named DRONE [26]–[30] to perform a functional mapping between the measured data and the underlying tissue parameters. In this work, we extended the DRONE approach to enable reconstruction of the high-dimensional CEST-MRF signals in a clinical setting. The benefits of DRONE include nearly instantaneous reconstruction time and elimination of the requirement for large patient-derived training datasets. The neural network used in this work was also trained on simulated data, as in the original DRONE method, but in this case training data was generated by solving the Bloch-McConnell equations for a 3 pool (water, solute, semi-solid) model.

### 2.2.1. Training dataset generation

The neural network was implemented in PyTorch [31] and consisted of a 30-node input layer (corresponding to the 30-point magnitude images acquired with the CEST-MRF sequence), two fully connected hidden layers with 300 nodes per layer and a 6-node output layer (Figure 1D). The output layer corresponded to the six parameters measured by the CEST-MRF sequence consisting of water T1 relaxation (T1w), water T2 relaxation (T2w), amide exchange rate (ksw), amide volume fraction (fs), semi-solid exchange rate (kssw) and semi-solid volume fraction (fss). A training dataset was generated by sampling the tissue parameter ranges using latin hypercube sampling [32] and simulating a CEST-MRF acquisition by solving the Bloch-McConnell equations. All signals were normalized to have a l2-norm of 1. The transmit field inhomogeneity (B1) was included in the training dictionary but not in the error calculation to induce the network to minimize the error in the other parameters instead. To accelerate the training set generation the CEST-MRF simulation was implemented on a Nvidia RTX2080 Ti GPU (Nvidia Corp. Santa Clara, CA) with 11GB of memory which enabled parallel processing of the training set entries. A fraction (20%) of the dataset was used as a validation set to assess the quality of network training with the remainder (80%) used for training. The network was trained for 4000 epochs with the Adam optimizer [33] using an l1-norm loss with a batch size of 1000 and an adaptive learning rate with weight decay of $10^{-4}$. Zero-mean Gaussian noise with 1% standard deviation was added to the training dictionary to promote robust learning. Network training required approximately 30 minutes on the GPU whereas quantification of the six parameter maps with the trained network required only ~100 ms for an image with 256×256 voxels.

## 2.3. Numerical simulations

The feasibility of DRONE reconstruction for six parameter maps was assessed in a custom modified Brainweb-based [34] digital phantom. The segmented grey matter (GM), white matter (WM) and cerebrospinal fluid (CSF) phantom maps were used to assign quantitative values, representative of the healthy brain, for each tissue type and parameter (Supporting Information Table S1). The digital phantom was used to simulate a CEST-MRF acquisition with the sequence and acquisition schedule described in section 2.1. The simulated data was reconstructed using a DRONE network trained with a training set of 60,000 entries sampled from the ranges shown in Supporting Table S2. The error between the reconstructed tissue parameter and the reference values was calculated as Error = 100×|Reference – Reconstructed|/Reference.

White Gaussian noise was added to the simulated data to study the effect of noise on the reconstruction for varying levels of signal-to-noise ratio (SNR). The SNR was defined as $20 \cdot \log_{10}(S/N)$ where S was the average white matter signal intensity for the acquisition and N was the noise standard deviation. The SNR was varied from 20 to 80 dB in intervals of 5 dB [35] and the data reconstructed with the same network for each SNR level. The normalized RMS error (NRMSE) was used to calculate the error between the estimated and reference values for the different SNR levels.

### 2.4. In vivo studies

All experiments were conducted on a 3T GE Signa Premier (GE Healthcare, Waukesha, WI) with the built-in transmit body coil and a 48-channels head coil for reception.

#### 2.4.1. Healthy volunteer subject

A healthy 34 years old female volunteer was recruited for this study and provided informed consent in accordance with our institution's IRB protocol. The subject was scanned with the CEST-MRF sequence described in section 2.1 with the following image acquisition parameters: field of view (FOV)= 280×280 mm$^2$, matrix size=256×256, in-plane resolution=1.1×1.1 mm$^2$, slice thickness=5 mm, echo time (TE) = 24 ms, partial Fourier, number of averages (NEX) = 1, TR = 3500 ms, FA =90°, $T_{sat}$ = 2560 ms, bandwidth = 250 kHz. The total scan time was 105 seconds. The measured data was reconstructed with the same network defined in section 2.3. The SNR for the in vivo acquisition, calculated as described in section 2.3, was approximately 53 dB.

##### 2.4.1.1. Sensitivity to B0 inhomogeneity

A B0 map was acquired using a dual-echo gradient echo sequence to assess the impact of B0 inhomogeneities on the CEST-MRF parameters. Seven regions-of-interest (ROI) were selected corresponding to areas with B0 values in a 50 Hz range. The correlation between the tissue parameter values and the B0 values in the ROIs was calculated for each parameter.

##### 2.4.1.2. In vivo reproducibility

The in vivo reproducibility of the CEST-MRF sequence was assessed by test-retest scanning. The healthy volunteer was sequentially scanned twice with the CEST-MRF sequence and then removed from the scanner. Following a 5 minutes delay, the subject was again placed in the scanner, re-localized and scanned two more times with the CEST-MRF sequence. Data from each scan were reconstructed with the trained DRONE network and gray matter (GM) and white matter (WM) ROIs defined. The ROIs were then used to determine the GM and WM values for all parameter maps. The Lin's concordance correlation coefficient (CCC) [36] was calculated in the GM and WM of each tissue parameter as a measure of the reproducibility of each parameter.

### 2.4.1.3. *Comparison with conventional MRF derived T1 and T2 maps*

The water T1 and T2 relaxation maps obtained with CEST-MRF were compared to an optimized conventional MRF-EPI sequence [26], [35]. The imaging parameters were kept the same as the CEST-MRF acquisition with the exception of the initial adiabatic inversion pulse with inversion time of 50 ms which preceded the data acquisition [20]. The acquired data were processed by a separate DRONE network and the total acquisition time for the optimized 50-point schedule was approximately 6 seconds per slice. The GM/WM ROIs defined in section 2.4.1.2 were used to calculate the reference mean±SD T1 and T2 values.

### 2.4.1.4. *Comparison with conventional CEST imaging*

Validation of the CEST-MRF parameters is difficult because no reference method is available. Instead, CEST spectra were synthesized from the CEST-MRF parameters and compared to experimentally measured spectra in the same subject. The subject was scanned with a CEST sequence with 57 resonance frequency offsets in the -7.2 to 7.2 ppm range. The FA, $B1_{sat}$, TR and $T_{sat}$ were set to 90°, 2 µT, 8 and 3.04 seconds respectively. The total scan length for the CEST acquisition was 7.6 minutes. The same acquisition parameters were used along with the CEST-MRF derived parameters as inputs to a Bloch-McConnell equation simulator to generate the synthetic spectra and the Pearson correlation between the two curves calculated.

### 2.4.1.5. *Comparison with dictionary fitting*

The performance of the DRONE network for the in vivo data was compared to standard dot-product pattern matching with a dictionary composed of 4 million entries generated using the parameter ranges listed in Supporting Table 2. Dictionary generation and matching required approximately 4 hours on a desktop server with 256 GB of memory.

### 2.4.2. **Patients with brain metastases**

Four patients with brain metastases were recruited for this study and gave informed IRB consent. The patients were scanned per the consensus standardized brain tumor imaging protocol [37] and institutional standard-of-care that included T1-weighted sequences pre- and post-gadolinium (Gd) contrast injection, FLAIR, diffusion, and perfusion acquisitions. All patients were previously treated with radiation therapy and/or chemotherapy. The proposed CEST-MRF was acquired prior to Gd injection using the same imaging parameters and tissue quantification as for the healthy volunteer scans (section 2.4.1) other than the resolution which was set to 1×1×3 mm$^3$ to enable visualization of small lesions. The tumor was contoured into ROIs comprising a necrotic core, solid core and edema as well as a contra-lateral healthy region by one of the authors (KRT), a trained radiation oncologist. Statistical significance of differences in the reconstructed tissue map values for each ROI in a representative patient (patient 1) was evaluated using a multi-comparison ANOVA test with Tukey HSD [38] with a significance level set at P=0.05.

## 3. Results

### 3.1. Numerical simulations

The DRONE reconstruction of the 6 tissue parameter maps in the simulated digital phantom is shown in Figure 2 for a high SNR simulation (80 dB) to isolate the intrinsic error in the DRONE reconstruction for the highly under-determined CEST-MRF data. The impact of SNR on the quantitative accuracy for each parameter is shown in Figure 3 on a log-log scale. Despite the high dimensionality of the data and the small training dictionary, the DRONE reconstruction accurately quantified the tissue maps yielding a NRMSE of less than 7% for all the parameters. As expected, increasing SNR reduced the NRMSE for all tissue parameters although the rate of improvement varied for each parameter which is reflective of the intrinsic sensitivity of the sequence to the different parameters. As an example, the NRMSE at 55 dB was approximately 8% for fss but 3% and 6% for T1w and T2w.

### 3.2. Healthy human volunteer

The quantitative CEST-MRF tissue parameter maps for the healthy human subject are shown in Figure 4A. The mean and standard deviation of the GM and WM values for each tissue parameter are listed in Table 1 along with the reference water T1 and T2 values obtained with the conventional MRF sequence which are shown in Figure 4B. The mean B0 value across the slice was approximately 4 Hz but varied significantly (approximately Δ0.5 ppm range) near the frontal sinuses and temporal regions, as expected. The tissue parameter changes for varying B0 values are shown in Figure 5A-F. All parameters were essentially uncorrelated with B0 ($R^2 <$ 0.08) except for T2w which was only poorly correlated ($R^2=0.33$). This is also evident in the tissue maps of Figure 4 where the susceptibility differences near the sinuses and the EPI readout used gave rise to geometrical distortions apparent in that region but little variations in the CEST-MRF parameters. The dictionary fitted data (Supporting Figure S1) yielded tissue maps that were heavily discretized (due to the fixed increment used) and noisier compared to the DRONE output.

### 3.2.1. In vivo reproducibility

The mean tissue parameter values for each of the 4 repeated scans are shown for GM and WM in Figure 6 and the corresponding CCC values are tabulated in Table 1. The repeated scans showed strong reproducibility (CCC > 0.93) for all parameters despite uncorrected registration errors induced by patient motion between scans.

### 3.2.2. Comparison with conventional CEST

The synthesized CEST curves for representative GM and WM points are shown overlaid on the measured CEST curves in Figure 7. There was overall good agreement between the measured and synthesized curves with a Pearson correlation of 0.98 for both GM and WM. Differences in the curves in the negative offset regions were due to nuclear Overhauser effects that were not included in the CEST-MRF model.

*Table 1: Estimated CEST-MRF parameter values and concordance correlation coefficient (CCC) for GM and WM in a healthy volunteer.*

|  |  | This study | | | | Conventional MRF | Literature | Reference |
|---|---|---|---|---|---|---|---|---|
|  |  | **Mean±SD** | **95% CI** | **CCC** | **95% CI** | **Mean±SD** | **Mean±SD** |  |
| *T1w (ms)* | WM | 864±166 | [847, 881] | 0.9717 | [0.9713, 0.972] | 821±35 | 956±217[a] | Bojorquez et al. [39] |
|  | GM | 1403±203 | [1383, 1423] | 0.98399 | [0.98378, 0.984] | 1527±429 | 1482±150[a] |  |
| *T2w (ms)* | WM | 74.0±5.31 | [73.4, 74.5] | 0.99501 | [0.99494, 0.9951] | 77.8±7.20 | 75±3[b] | Lu et al. [40] |
|  | GM | 80.6±8.69 | [79.8, 81.5] | 0.99403 | [0.99394, 0.99411] | 91.7±28.1 | 83±4[b] |  |
| *ksw (Hz)* | WM | 48.4±5.09 | [47.9, 48.9] | 0.99214 | [0.99203, 0.9923] |  | 42.3±2.9 | Perlman et al. [41] |
|  | GM | 50.0±3.84 | [49.6, 50.4] | 0.99168 | [0.99157, 0.99179] |  | 34.6±9.5 |  |
| *kssw (Hz)* | WM | 23.6±5.92 | [23.0, 24.2] | 0.9645 | [0.96397, 0.96502] |  | 23±4 | Stansiz et al. [42] |
|  | GM | 14.3±4.66 | [13.8, 14.7] | 0.96127 | [0.96069, 0.96183] |  | 40.0±1 |  |
| *fs (%)* | WM | 0.504±0.098 | [0.491, 0.513] | 0.96436 | [0.96386, 0.96486] |  | 0.31±0.02 | Perlman et al. [41] |
|  | GM | 0.601±0.057 | [0.595, 0.607] | 0.98506 | [0.98484, 0.98528] |  | 0.32±0.07 |  |
| *fss (%)* | WM | 9.83±0.943 | [9.74, 9.93] | 0.98786 | [0.98768, 0.98804] |  | 8.9±0.3 | Stansiz et al. [42] |
|  | GM | 6.19±1.78 | [6.01, 6.36] | 0.93658 | [0.93572, 0.93744] |  | 4.4±0.4 |  |

[a] mean of all reported values
[b] mean of reported occipital and frontal gray matter

### 3.3. Patients with brain metastases

The CEST-MRF parameter maps for a representative patient are shown in Figure 8A along with the T1-weighted (pre-, post-Gd), FLAIR, diffusion and perfusion scans from the standard brain protocol for comparison. Box and whiskers plots of the CEST-MRF tissue map values for each ROI in the same patient are shown in Figure 8B with the median and first and third quartile ranges. The differences between the parameter values in the different ROIs for all tissue parameters were statistically significant (P=0.05) except as shown. A box and whiskers plot for all patients is shown in Figure 9. There were notable trends in the parameter maps. The T1w and T2w in the tumor were both elevated in comparison to the contra-lateral tissue, particularly in the necrotic region. The amide exchange rate in the necrotic region was lower which is suggestive of a lower pH environment. The amide volume fraction was also reduced in the necrotic and edema regions.

## 4. Discussion

This is the first study to demonstrate the feasibility and utility of CEST-MRF in clinical cancer imaging. In addition to a short acquisition using an EPI readout, the combination of CEST-MRF with deep learning quantification using DRONE enables nearly immediate parameter quantification at minimal computational cost, which would potentially allow for image evaluation at the scanner and facilitate clinical adoption.

### 4.1. Effect of schedule on quantitative parameters

The CEST-MRF signal depends on multiple acquisition parameters including the shape of the saturation pulses, duration and frequency, the excitation flip angle, saturation power, and others. The dependence on multiple parameters can be beneficial in improving the discrimination between signals from different tissues. The semi-solid pool can particularly benefit from inclusion of additional resonance frequency offsets in the schedule given the very broad linewidth of the MT pool which would improve kssw discrimination. For simplicity, in this work only the saturation power was varied using a random schedule that is unlikely to be optimal. The numerical phantom results (Figure 3) exemplify this since increasing the SNR reduced the NRMSE for all tissue parameters, as expected, but the rate of improvement varied by parameter. This reflects the intrinsic sensitivity of the sequence to each parameter which can be optimized by modifying the acquisition schedule. Indeed, work by our group and others has shown that simultaneously varying multiple acquisition parameters and optimizing the acquisition schedule can markedly improve tissue discrimination, reduce the sensitivity to noise and shorten the total scan time [25], [35], [43]–[45]. However, the optimization of CEST-MRF schedules is challenging because of the large number of parameters (high dimensionality) of the optimization problem. To overcome this difficulty, we have previously introduced a deep learning schedule optimization approach for CEST-MRF optimization and demonstrated it on a preclinical scanner [24], [46]. This method can be readily adapted for the clinical CEST-MRF sequence described in this work and is expected to significantly improve the sensitivity to noise and the accuracy of the tissue map quantification. Future work will explore this idea.

### 4.2. Neural network reconstruction of high dimensional signals

While the original DRONE was only applied for T1 and T2 mapping [26], the method is capable of simultaneous estimation of a much larger set of parameters [28]–[30], [41]. Nevertheless, there are important challenges associated with reconstruction of high dimensional signals. First, for the network to correctly estimate the underlying tissue parameters, the training set must adequately cover the parameter space. Unfortunately, due to the "curse of dimensionality", this requires large training datasets and consequently long processing time. To overcome this problem, we used a regular sampling of the parameter space and implemented the training dataset generation on a GPU to parallelize the processing. This enabled the use of a small (60,000 entries) 7-dimensional dictionary comprising of the 6 tissue parameters and the instrumental parameter B1. Although the B1 was included in the training dictionary, to avoid the risk of the network converging to spurious solutions given the high dimensionality of the problem, B1 was excluded from the network output and the training error calculation. The network training therefore minimized errors in the tissue parameters alone while still accounting for the inevitable B1 variations in vivo. Since T1 estimation is biased by B1, the inclusion of B1 mitigates the T1 underestimation described in prior CEST-MRF studies [15]. Although B0 was not included in the network training, there was only poor correlation between the B0 values and the tissue parameters and little variation in the tissue parameter values across the measured slice (Figure 4), despite significant susceptibility differences near the sinuses. This illustrates the intrinsic robustness to B0 inhomogeneity of the CEST-MRF pulse sequence combined with the DRONE reconstruction. The insensitivity of the CEST-MRF method to B0 shifts results from the use of a fixed saturation frequency offset and the normalization of the trajectory by the trajectory norm. As long as the saturation pulse is spectrally localized within the relatively broad amide proton resonance, shifts in the irradiation position caused by B0 field inhomogeneity will only result in a different scaling of the signal trajectory for a given voxel, which will be normalized out when taking the norm of the trajectory. However, more severe B0 inhomogeneity that may be encountered in future 3D whole brain studies may need to be addressed by including B0 in the training dictionary as was done for B1.

### 4.3. In vivo studies

There was generally good agreement between the tissue parameter values obtained with the CEST-MRF sequence and alternative methods. While there is a wide range of reported values in the literature, the mean WM and GM T1w/T2w measured in this study (Table 1) agreed with both the reference conventional MRF values as well as literature values from multiple studies using a variety of different quantification techniques [39]. Optimizing the schedule to improve the T1 discrimination (by additionally varying FA and TR, for instance) and refining the network's B1 estimation may address this issue but further study is needed to confirm this. The mean amide exchange rates measured in this study (WM/GM=48.4±5.09 / 50.0±3.84 Hz) were similar to that measured in vivo with CEST-MRF in preclinical [15], [41] and clinical [23] models and with spectroscopic methods [19]. There was a clear delineation between GM and WM in the fss map which is expected given the high WM myelin content and the sensitivity of the semi-solid volume fraction to lipid content. In general, validating the CEST-MRF parameters is difficult because no gold-standard exists for all parameters. As an alternative, a CEST spectrum synthesized from the CEST-MRF parameters was compared to a measured CEST spectrum in vivo. The agreement between the two spectra (r=0.98) is indicative of the accuracy of the reconstructed CEST-MRF maps though it should be noted that there were some differences between the measured and synthetic spectra in the negative offset region of the spectrum. This is understandable given that nuclear Overhauser effects, arising from aliphatic protons with chemical shifts between -2 and -5 ppm, were not included in the CEST-MRF model but are present in the experimentally measured CEST spectrum. The test-retest experiments (Figure 6, Table 1) demonstrate the strong reproducibility of the CEST-MRF tissue maps, despite potential uncorrected subject motion.

Further work is necessary to confirm the reproducibility, which is a critical feature for longitudinal studies or treatment response monitoring applications for this technology.

### 4.4. Brain tumor studies

In many tumors, changes to the tissue parameters occur simultaneously and can therefore confound the CEST signal. In contrast, the combination of the CEST-MRF maps allows separation of different contributors to the signal which can improve tumor characterization. The significantly different values between the drawn tumor ROIs (Figure 8B, Figure 9) can improve tumor segmentation and may provide more specific tumor characterization. It should be noted that the lesion shown in Figure 8 is one that was previously irradiated which is sure to affect the CEST-MRF maps. This is also evident in the perfusion (Ktrans, vp) maps (Figure 8A) which are suggestive of a treated tumor. Some features in the CEST-MRF maps, like the reduced amide volume fraction (fs) in the necrotic and edema regions, can be understood as resulting from disrupted protein synthesis in necrotic cells and diluted protein concentrations due to edema. Similarly, the decreased amide exchange rate in the necrotic core is consistent with a decreased pH as expected for apoptotic and necrotic tumor regions [47]–[49]. Finally, the reduced semi-solid volume fraction (fss) in the lesion can be a result of demyelination in that region or post-treatment effects. At present, the biological effects of radiation on the CEST-MRF tissue maps are not well understood and represent an additional confounding factor. Furthermore, as demonstrated by the results from multiple patients (Figure 9), brain metastases arising from different primary tumor histologies showed varying parameter values. Because of the intrinsic biological variability in tumors, prospective large-scale studies will be required to draw meaningful conclusions about relationships between the CEST-MRF maps and the associated tumor characteristics.

### 4.5. Limitations and future work

In the current version of the sequence, coverage is limited to a single slice which may be inadequate for tumors with large spatial extent. Incorporating simultaneous multi-slice and/or slice interleaving can resolve this issue without increasing the total scan length. An optimized ordering of the slice interleaving can also improve tissue discrimination as previously shown [50]. The availability of a rapid and non-invasive method for imaging endogenous amide exchange rates and pH makes possible many different studies. One example is the imaging of the tumor's response to oncolytic virotherapy, as recently demonstrated in a preclinical model [27]. Such studies will facilitate development of personalized therapies and can help improve treatment outcomes.

## 5. Conclusion

This work demonstrated the feasibility of fast EPI acquisition and fast deep learning-based parameter reconstruction algorithm for quantitative CEST-MRF imaging of brain tumors on a clinical scanner. The proposed CEST-MRF presented good reproducibility and quantitative results that are consistent with conventional qualitative MRI. CEST-MRF can be particularly beneficial in complex pathologies such as brain tumors.

## 6. Conflict of interest

OC and CTF hold a patent on the CEST-MRF technology.

## 7. Acknowledgments

The authors are grateful to Luanna Chan and Tara Fahy for assistance with patient recruitment. This work was supported by NIH/NCI grant P30-CA008748. OP acknowledges funding from the European Union's Horizon 2020 research and innovation program under the Marie Skłodowska-Curie grant agreement No 836752 (OncoViroMRI). This paper reflects only the authors' view, and the Research Executive Agency of the European Commission is not responsible for any use that may be made of the information it contains.

## 9. Figure Captions

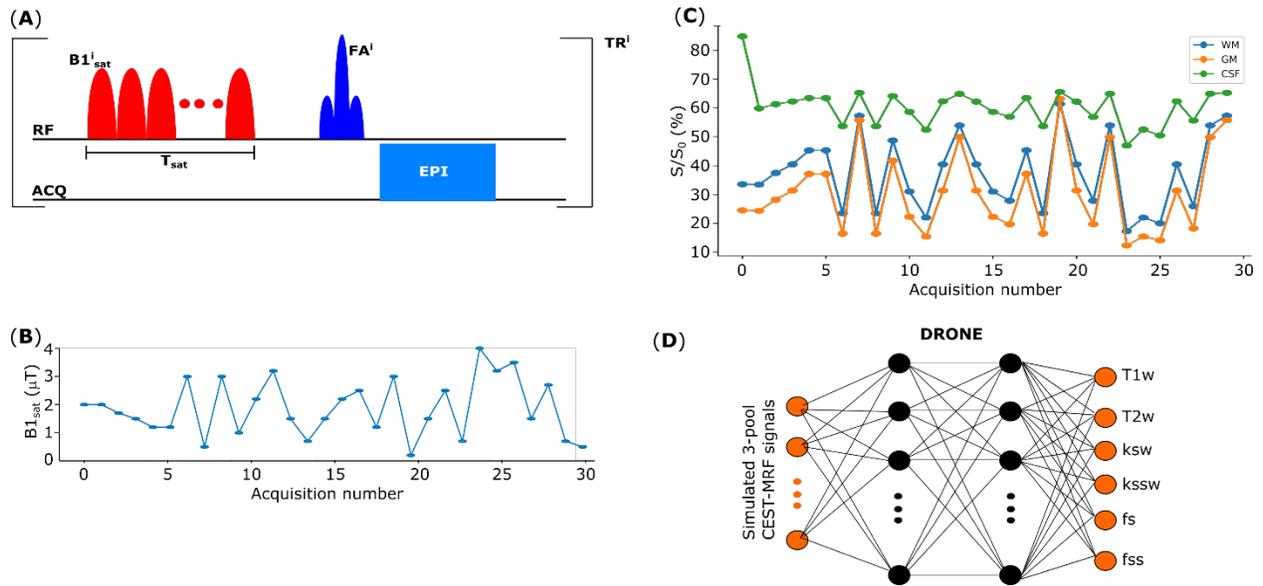

**Figure 1**: (A) Clinical CEST-MRF pulse sequence shown for one schedule point. The magnetization is saturated with a Gaussian-shaped pulse train and exchanges with the water. The water signal is then excited and read out with an EPI k-space sampling. The saturation pulse train power ($B1_{sat}$) and duration ($T_{sat}$) and the excitation pulse flip angle (FA) are varied according to the MRF acquisition schedule. For simplicity, only the saturation power was varied in this study. (B) Schedule of $B1_{sat}$ powers. (C) Sample CEST-MRF signals for GM, WM and CSF. (D) DRONE network used in quantification of the CEST-MRF data. The network is trained with simulated 3-pool data and outputs the water relaxation (T1w, T2w), amide exchange (ksw, fs) and semi-solid (kssw, fss) parameters.

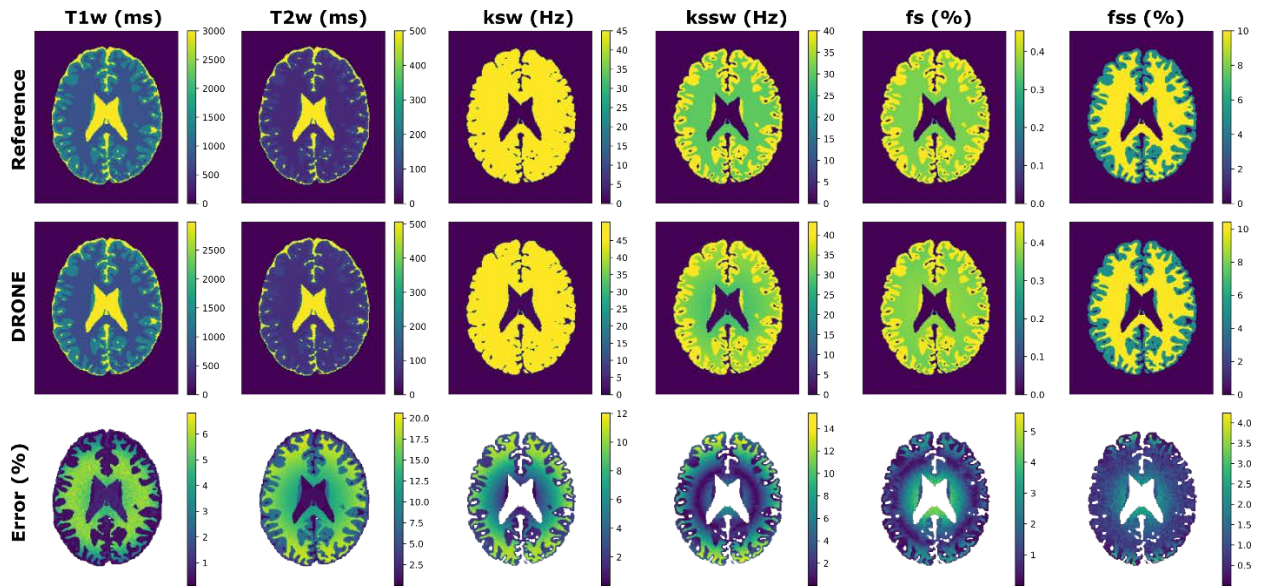

**Figure 2**: DRONE reconstruction of 6 parameters in a digital phantom in comparison to the reference values. Regions associated with the background, skull and scalp were set to zero. The error, calculated as 100 ×|Reference – DRONE|/Reference, is shown for each tissue. Note the effect of the B1 inhomogeneity visible in the error maps.

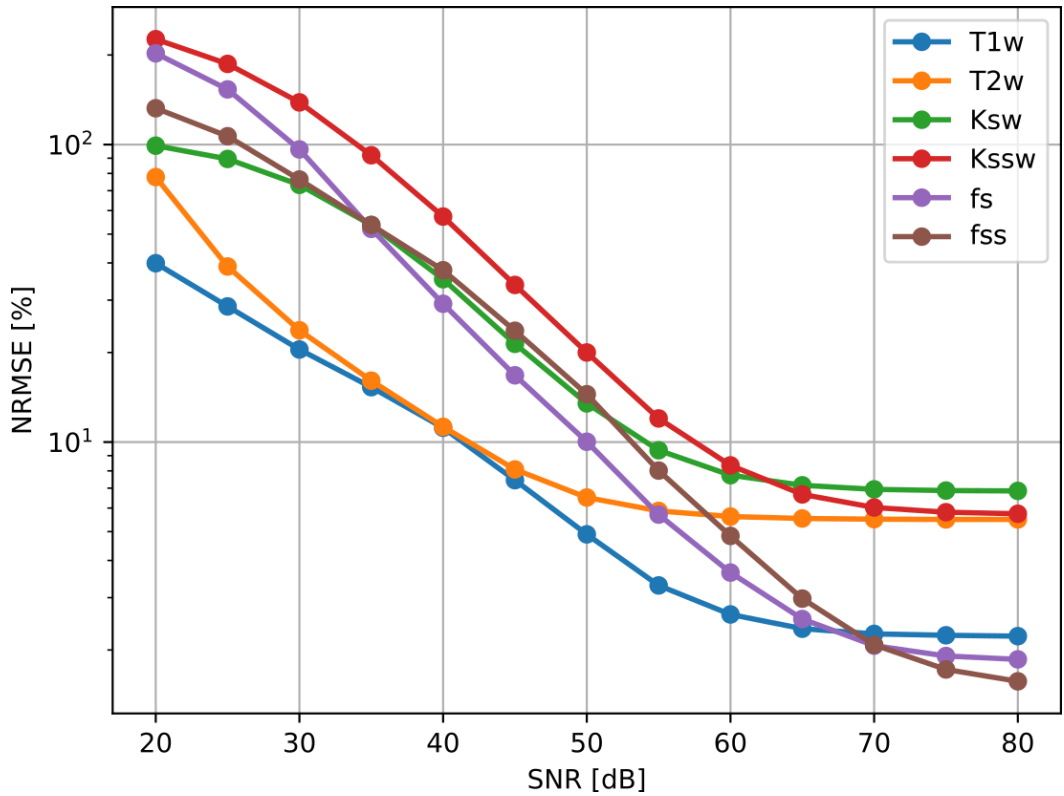

**Figure 3**: Normalized RMS error (NRMSE), on a log scale, of the DRONE reconstructed CEST-MRF maps in a digital brain phantom using a random schedule for varying levels of added white gaussian noise. Changes in SNR non-linearly affected the NRMSE of the different parameters illustrating the sensitivity of the sequence and schedule to each tissue parameter.

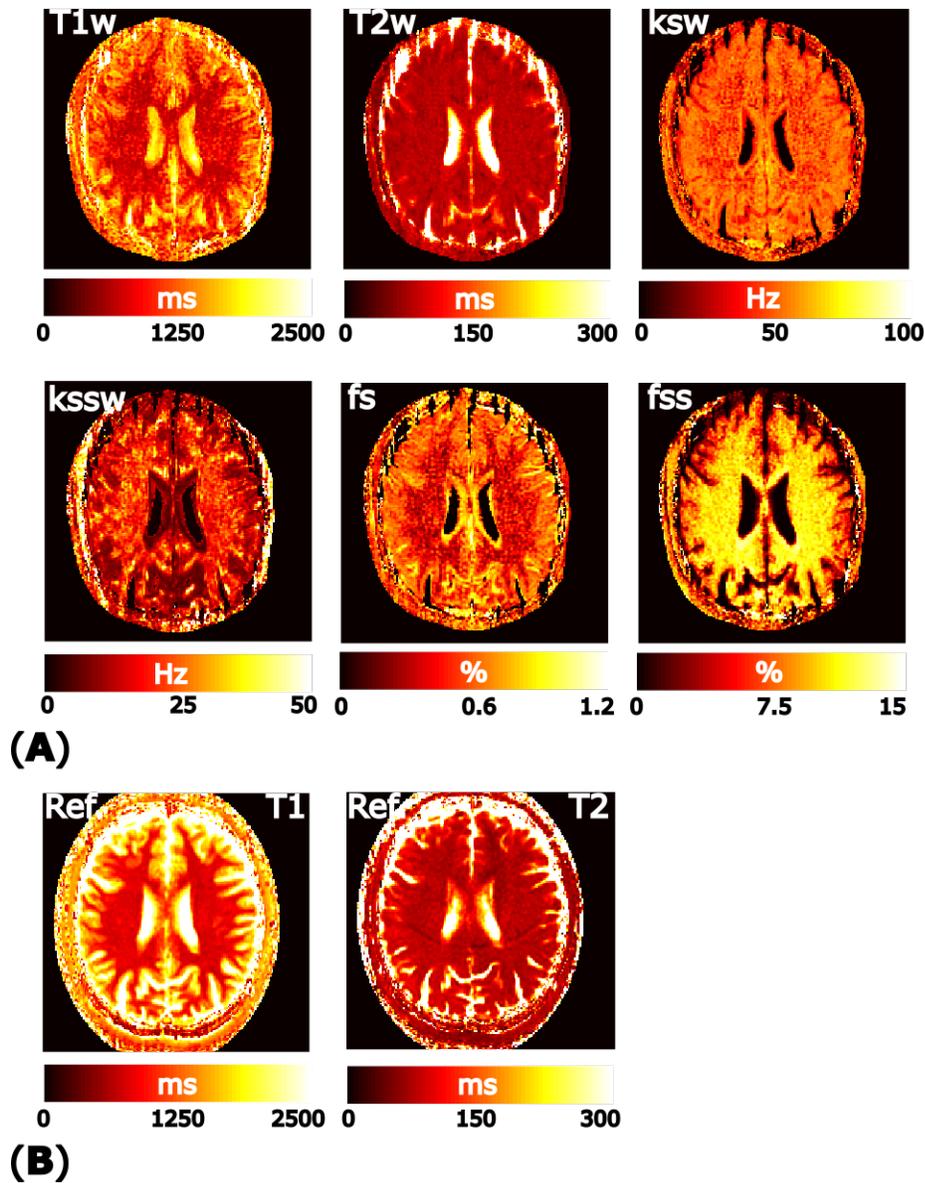

**Figure 4**: (A) Reconstructed tissue parameter maps obtained from a healthy volunteer with the CEST-MRF method. Note the elevated semi-solid volume fraction in the WM reflective of the higher myelin content. (B) Reference T1 and T2 maps obtained with the optimized MRF-EPI sequence.

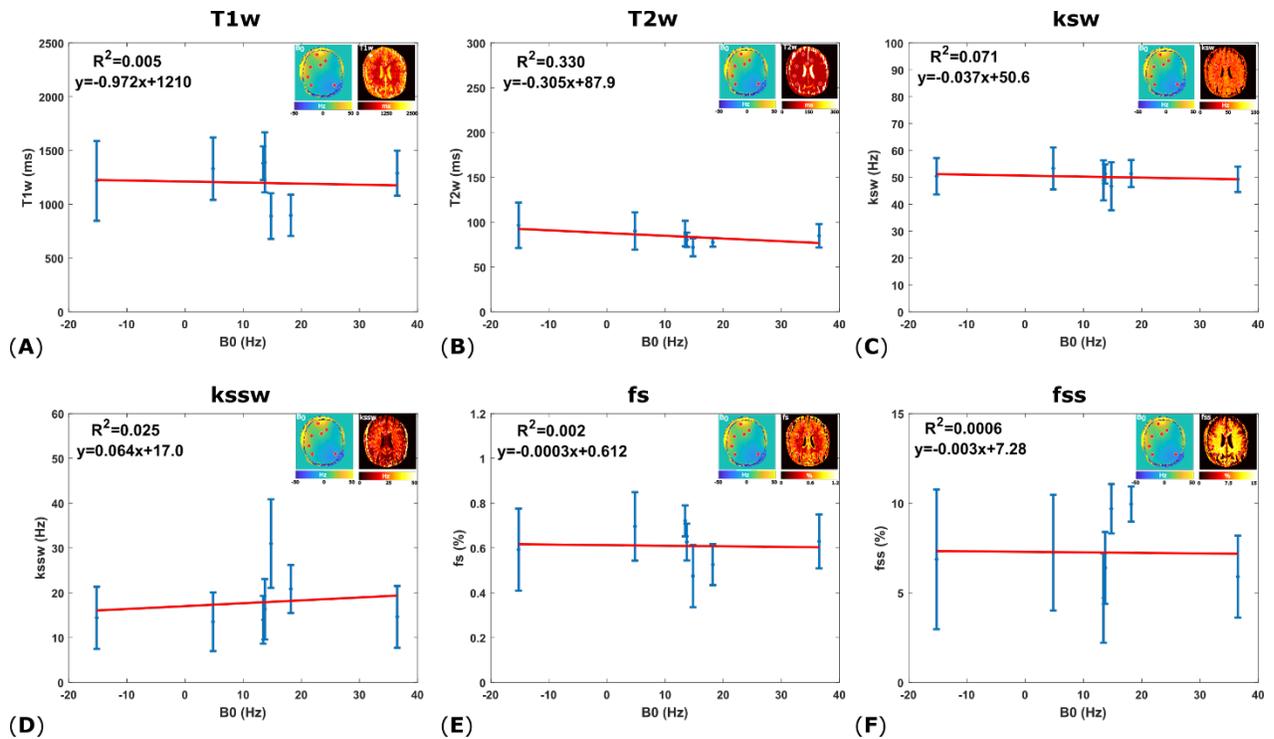

**Figure 5:** Variation in the tissue parameter maps as a function of the B0 inhomogeneity. Shown are the water relaxation parameters: (A) T1w, (B) T2w, amide parameters: (C) ksw, (E) fs and semi-solid parameters: (D) kssw, (F) fss. Note the poor correlation between the B0 values and the different parameters illustrative of the robustness of the sequence to B0 variations.

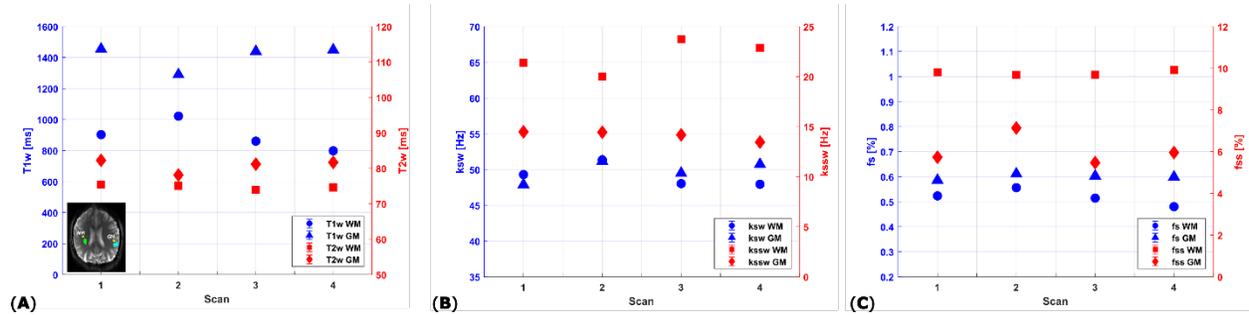

**Figure 6**: In vivo GM and WM tissue parameter values for the four CEST-MRF scans in the healthy volunteer. Scans 1 and 2 were acquired in the first session and scans 3 and 4 in the second session. Blue entries correspond to the left y-axis and red entries to the right y-axis with error bars omitted for clarity. (A) T1w and T2w, (B) ksw and kssw, (C) fs and fss. The locations of the WM and GM regions used are shown inset in (A). Note the good repeatability between scans. The concordance correlation coefficient for each parameter and tissue type is listed in Table 1.

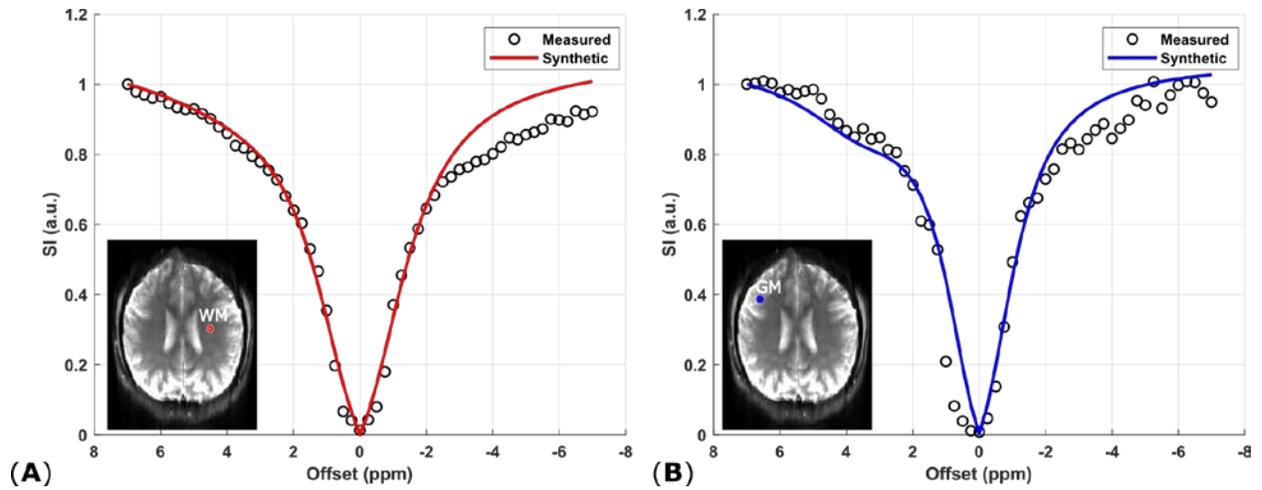

**Figure 7:** Comparison between a measured CEST spectrum and one synthesized from the CEST-MRF parameters for (A) WM and (B) GM. Nuclear Overhauser effects, not included in the CEST-MRF model, led to the discrepancy between the curves in the negative offsets' region. The measured and synthetic curves were nevertheless highly correlated (r=0.98).

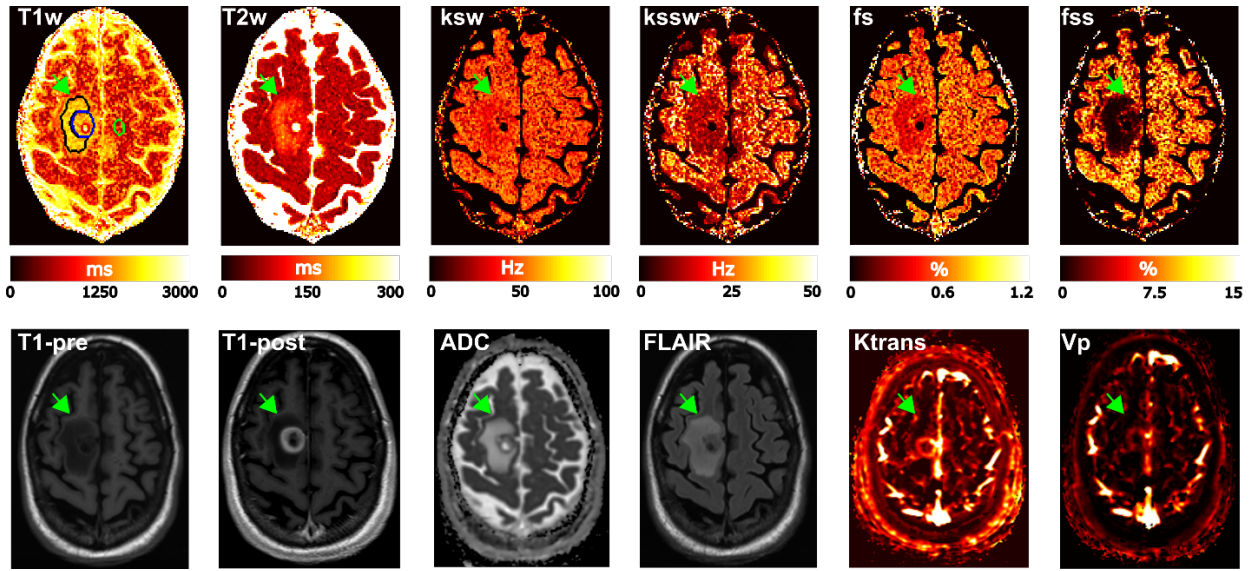

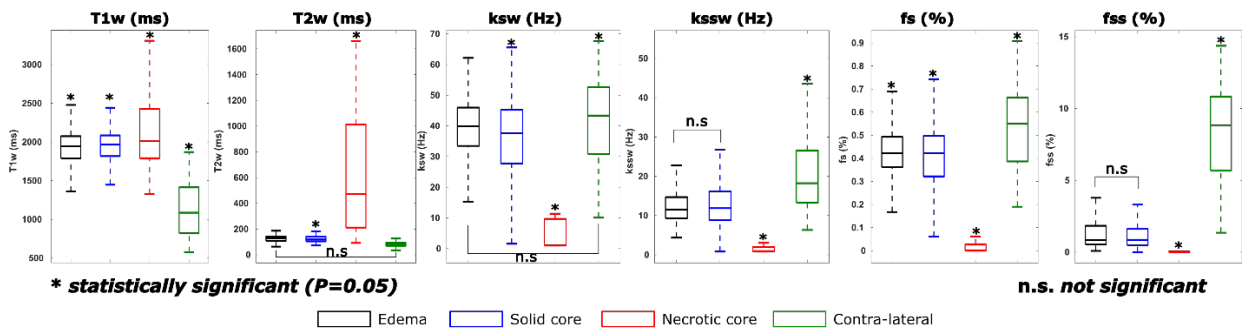

**Figure 8**: (A) In vivo CEST-MRF maps from a patient with brain metastasis and the corresponding images from a standard clinical protocol for comparison. Green arrows indicate the location of the lesion. The segmented tumor regions are denoted by the colored outlines on the T1w map and include the edema (black), solid core (blue), necrotic core (red) and contra-lateral (green) regions. The T1-pre and T1-post denote the T1-weighted acquisition before and after contrast injection whereas Ktrans and Vp refer to the perfusion and plasma volume maps. The marked differences in the tissue map values between the lesion and healthy tissues are notable. (B) Box and whiskers plots of the reconstructed tissue maps values for the different ROIs. The distribution of the parameter values along with the median and the first and third quartile ranges are shown. The differences between the tumor ROIs were all statistically significant (P=0.05) as determined by a multi-comparison ANOVA test with Tukey HSD except the regions denoted in the figure by "n.s".

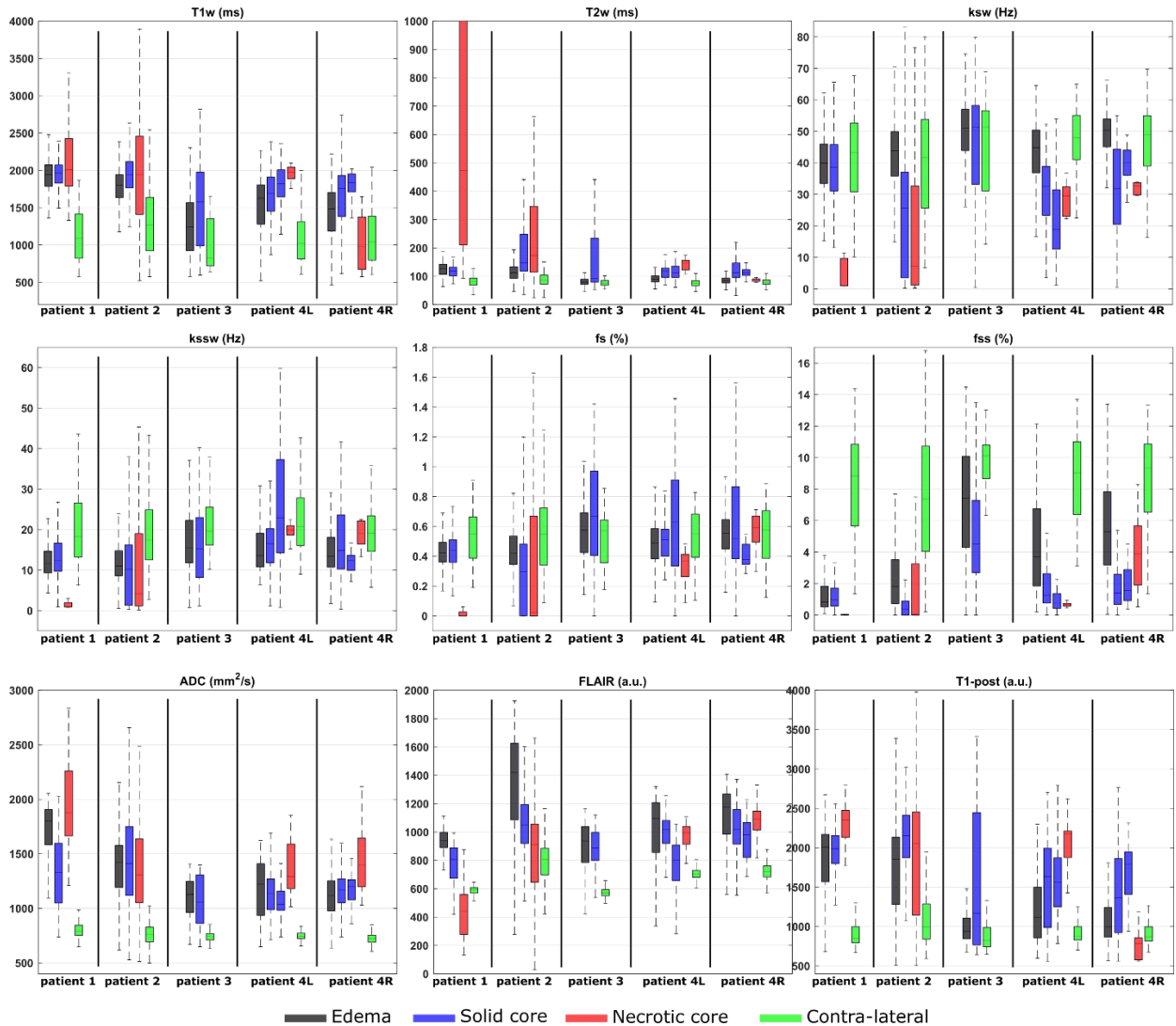

**Figure 9**: Box and whiskers plots for CEST-MRF-derived T1, T2, ksw, kssw, fs, and fss as well as conventional sequence values from T1 post Gd contrast, ADC, and FLAIR images. Patient #1: 61 year-old male, NSCLC adenocarcinoma, Patient #2: 65 year-old female, melanoma, Patient #3: 48 year-old female NSCLC adenocarcinoma/small cell, Patient #4: 40 year-old male, melanoma (L: left side with two tumors and surrounding region. R: right side with two tumors and surrounding regions).

# Supporting Information

**Supporting Table S1:** Tissue parameter values of the digital phantom used.

**Supporting Table S2**: Tissue parameter ranges used in the training dataset and dictionary generation. Dictionary parameters bounds are formatted as lower:interval:upper.

**Supporting Figure S1**: Dictionary matched reconstruction of the in vivo healthy subject data. The ranges used to generate the 4 million dictionary entries are shown in Supporting Table S2. While large, the dictionary size used (4 million entries), is insufficient to cover the 7-dimensional space of tissue parameters leading to the heavily discretized and noisy appearance of the maps, unlike the equivalent DRONE reconstruction shown in Figure 4.